\pdfoutput=1

\documentclass[aps,prl,10pt,twocolumn,showpacs,superscriptaddress]{revtex4-1}
\usepackage{amsmath}
\usepackage{latexsym}
\usepackage{amssymb}
\usepackage{bm}
\usepackage{graphics, epstopdf}
\usepackage{color}

\usepackage{newlfont}
\usepackage{amssymb}
\usepackage{amsfonts}
\usepackage{amsmath}
\usepackage{amsthm}
\usepackage{graphicx}
\usepackage{epsfig}
\usepackage{color}
\usepackage{bm}

\newcommand{\ket}[1]{|{#1}\rangle}

\usepackage{times}

\begin{document}
\title{Super Quantum Discord with Weak Measurements}

\author{Uttam Singh}
\author{Arun Kumar Pati}

\affiliation{Quantum Information and Computation Group\\
Harish-Chandra Research Institute, Chhatnag Road, Jhunsi, 
Allahabad 211 019, India}


\date{\today}

\begin{abstract}
Weak measurements cause small change to quantum states, thereby opening up the possibility of new ways of 
manipulating and controlling quantum systems.  We ask, can weak measurements reveal more 
quantum correlation in a composite quantum state? We prove that the weak measurement induced quantum discord, 
called as the ``super quantum discord'', is always 
larger than the quantum discord captured by the strong measurement. 
Moreover, we prove the monotonicity of the super quantum discord as a function of the measurement strength.
We find that unlike the normal quantum discord, for pure entangled states, 
the super quantum discord can exceed the quantum entanglement.
Our result shows that the notion of quantum correlation is not only observer dependent but also depends on how weakly 
one perturbs the composite system.


\end{abstract}

\maketitle

\emph{Introduction.}--
Quantum states are fragile to quantum measurements. Yet they try to maintain their privacy. This is exemplified by 
the fact that it is impossible to know the state given a single quantum system. 
If we measure an arbitrary quantum state in some orthogonal basis (projective measurement), we loose its coherence.  
However, if we perform measurement which couples the system and the measuring device weakly, then the system will be perturbed gently 
and may not loose its coherence completely. Such a scheme was proposed by Aharonov-Albert-Vaidman \cite{aav} which is called
weak measurement formalism. This gives a weak value of an observable which can take values outside the spectrum
of the observable. Remarkably, it has been shown that the weak measurements are universal in the sense that any generalized measurement can be realized as a 
sequence of weak measurements which result in small changes to 
the quantum state for all outcomes \cite{brun}.
The weak-value amplification has found several applications in recent years.
To name a few, weak measurements have been exploited for interrogating quantum systems in a coherent manner \cite{sandu}, 
in understanding the role of the uncertainty principle in the double-slit experiment \cite{wise1,mir}, in understanding the macrorealism \cite{will,pala}, in resolving 
Hardy's nonlocality \cite{js}. 
On the practical applications, this has been used for observation of photonic version of the spin Hall effect \cite{hos}, to 
amplify the deflection of optical beam in the Sagnac interferometer \cite{dix},
in the feedback control of quantum systems \cite{ga} and even direct measurement of wavefunction of single photon \cite{jsl}.
Recently, it is shown that weak measurements also help in protecting quantum entanglement from decoherence \cite{kim}.

Understanding the nature of correlations in composite systems is one of the prime goals in the emerging area of 
quantum information theory. For bipartite states, an entropic measure of quantum correlation known as the quantum discord has 
been proposed \cite{zurek}. The quantum discord represents the inaccessible information that cannot be extracted by measurements on one subsystem. 
It is the difference between the total and the classical correlation \cite{zurek,hend}. 
It turns out that unlike the entanglement, the quantum discord can be nonzero even for some separable states.
This suggests that quantum discord may capture quantum correlation for mixed states that goes beyond the entanglement.
The quantum discord has been investigated in a wider context starting from the possibility of giving the power to quantum 
computation in the absence of entanglement \cite{animesh}, quantum communication such as quantum state merging \cite{madhok,caval}, 
finding conditions for the monogamy nature \cite{prabhu,gio}, quantum entanglement distribution with separable states \cite{alex,chuan} and has 
been also experimentally used as a resource \cite{dak,gu}.
Recently, it has been shown that the quantum discord is also a physical quantity, because erasure of quantum 
correlation must lead to entropy production in the system and the environment \cite{akp}. (For a recent review on quantum discord 
see \cite{mera,modi}.)


Weak measurements are not only important in exploring fundamental physics questions but also have technological implications. 
In this paper, we ask, 
can they reveal more quantum correlation for a bipartite quantum system? If they can, then one can exploit this extra quantum 
correlation for information processing. 
We prove that the weak measurement 
performed on one of the subsystems can lead to `` super quantum discord'' (SQD) that is always larger than the normal quantum discord captured by the 
strong (projective) measurements. 
Furthermore, we prove that the super quantum discord is a monotonic
function of measurement strength and it covers all the values between the mutual information and the normal quantum discord.
Since quantum discord is regarded as a resource  
our result shows that the super quantum discord can be potentially a more useful resource. We illustrate the notion 
of the weak measurement induced discord for pure and mixed entangled states. A remarkable feature of the super quantum discord is that for pure entangled states
it can exceed the quantum entanglement. In this sense, SQD reveals quantum correlation that truly goes beyond
quantum entanglement even for pure states. Thus, quantum correlations are not only observer dependent 
but also depend on how gently or strongly one perturbs the quantum system. 
These findings will have several fundamental applications in quantum information processing 
and quantum technology.

{\it Weak measurements and super quantum discord}.--
The theory of weak measurements can be formulated using the pre- and post-selected quantum systems \cite{aav} 
as well as in terms of measurement operator formalism as
done by Oreshkov and Brun \cite{brun}. The later approach provides a new tool in quantum information theory to handle weak as well as strong measurements. 
Since a quantum measurement with any number of outcomes can be constructed as a 
sequence of measurements with two outcomes, one can consider dichotomic measurement operators. The weak measurement operators are given by
$P(\pm x) = \sqrt{\frac{(1\mp\tanh x)}{2}} \Pi_0 + \sqrt{ \frac{(1\pm\tanh x)}{2}} \Pi_1$,
where $x$ is a parameter that denotes the strength of the measurement process, $\Pi_0$ and $\Pi_1$ are two orthogonal projectors with 
$\Pi_0 + \Pi_1 =1$.
The weak measurement operators satisfy $P^{\dagger}(x)P(x) + P^{\dagger}(-x)P(-x) = 1$. These operators have the following properties:
(i) $P(0) = \frac{I}{\sqrt 2}$ resulting in no state change, (ii) in the strong measurement limit we have the projective measurement operators, 
i.e., $\lim_{x \rightarrow \infty} P(-x) = \Pi_0$ and  $\lim_{x \rightarrow \infty} P(x) = \Pi_1$, (iii) 
$ P(x) P(y)\propto P(x+y)$, and (iv) $[P(x), P(-x)] =0$.

Before defining the super quantum discord, let us recapitulate the normal quantum discord with the strong measurements. 
For a bipartite quantum state \(\rho_{AB}\), the total correlation is defined via the ``quantum mutual information''\cite{partovi,cerf}. 
It is given by 
$I(\rho_{AB})= S(\rho_A)+ S(\rho_B)- S(\rho_{AB})$,
where \(\rho_A\) and \(\rho_B\) are the local density matrices of \(\rho_{AB}\) and 
\(S(\sigma) = - \mbox{Tr} \left(\sigma \log_2 \sigma\right)\) is the von Neumann entropy of a quantum state \(\sigma\). 
The classical correlation is defined as \cite{hend}
$J_B(\rho_{AB}) = S(\rho_A) -S(\rho_{A|B})$,
where the ``quantum conditional entropy'' is given by  
$S(\rho_{A|B}) = \min_{\{\Pi_i^B\}} S(\rho_{A|\{\Pi^B \}}) = \min_{\{\Pi_i^B\}} \sum_i p_i S(\rho_{A|i})$,
with the minimization being over all projection-valued measurements, \(\{\Pi^B_i\}\),  performed on the subsystem \(B\).
The probability for obtaining outcome \(i\) is \(p_i = \mbox{Tr}_{AB}[(I_A \otimes \Pi^B_i ) \rho_{AB} ( {I}_A \otimes \Pi^B_i) ]\), and 
the corresponding post-measurement state  
for the subsystem \(A\) is \(\rho_{A|i} = \frac{1}{p_i} \mbox{Tr}_B[({I}_A \otimes \Pi^B_i) \rho_{AB} ({I}_A \otimes \Pi^B_i)] \), 
where \( {I}_A\) is the identity operator on the Hilbert space \( {\cal H}_A\).
The difference $[I(\rho_{AB})-J_B(\rho_{AB})]$ is a measure of quantum correlation, 
and called as the quantum discord \cite{zurek}. 
Therefore, the quantum discord with the strong measurement performed on the subsystem $B$ is given by 
\begin{equation}
\label{dis}
D(A:B)=  \min_{\{\Pi_i^B\}} \sum_i p_i S(\rho_{A|i}) - S(A|B),
\end{equation}
where $ S(A|B)= S(\rho_{AB}) - S(\rho_{B})$. 

Now, let us define what we call as the super quantum discord. It is defined as the quantum correlation in a 
state $\rho_{AB}$  as seen by an 
observer who performs a weak measurement on the subsystem $B$. 
In this case, the conditional entropy $S(A|\{\Pi^{B}\})$ is replaced by the weak quantum conditional entropy $S_w(A|\{P^{B}(x)\})$, 
where the subscript $w$ refers to the fact that one performs the weak measurement on the subsystem $B$ with the weak operators
$\{P^{B}(x)\}$. After the weak measurement, the post-measurement state for the subsystem $A$ is given by 
\begin{equation}
 \rho_{A|P^{B}(\pm x)}=\frac{\mbox{Tr}_{B}[(I \otimes P^{B}(\pm x)) \rho_{AB} (I \otimes  P^{B}(\pm x))]} 
{\mbox{Tr}_{AB}[(I \otimes P^{B}(\pm x)) \rho_{AB} (I \otimes P^{B}(\pm x))]}
\end{equation}
and the probability with which this occurs is given by 
\begin{equation}
 p(\pm x) = 
\mbox{Tr}_{AB}[(I \otimes P^{B}(\pm x)) \rho_{AB} (I \otimes P^{B}(\pm x))].
\end{equation}
The ``weak quantum conditional entropy'' is given by
\begin{equation}
 S_w(A|\{P^{B}(x)\})= p(x) S(\rho_{A|P^{B}(x)}) + p(-x) S(\rho_{A|P^{B}(-x)}).
\end{equation}
Note that the weak quantum conditional entropy $S_w(A|\{P^{B}( x)\}) = S_w(A|\{P^{B}(\{\Pi_i^B\}, x)\})$ is a function of the 
projectors and the measurement strength parameter $x$.
Therefore, the super quantum discord denoted by $D_w(A:B)$ is defined as
\begin{equation}
\label{def}
D_w(A:B):=  \min_{\{\Pi_i^B\}}  S_w(A|\{P^{B}(x)\}) - S(A|B).
\end{equation}
This is a positive quantity which follows from the monotonicity of the mutual information.
Next, we prove that the weak measurements can indeed reveal more quantum correlation of a bipartite quantum state, hence the 
name super quantum discord.

\noindent
{\bf Theorem 1}: Given a bipartite state $\rho_{AB}$, the super quantum discord (SQD) revealed by the weak measurement is always greater than or equal to 
the normal quantum discord with the strong measurement, i.e., $D_w(A:B) \ge D(A:B)$.

\noindent
{\bf Proof:} Consider the POVM elements for the weak measurement as given by $E(x) = P^{\dagger}(x)P(x)$ and $E(-x) = P^{\dagger}(-x)P(-x)$.
The strong measurement operators are given by $\{E_{0},E_{1}\}$, where $E_i=\Pi_i^{\dagger}\Pi_i=\Pi_i$ ($i=0,1$).
Now, we have 
$E(x) =\sum_{i=0}^{1}a_i(x)E_{i}$ and $ E(-x) = \sum_{i=0}^{1}a_i(-x)E_{i}$,
where $a_0(x)=\frac{(1-\tanh x)}{2}$ 
and $a_1(x)=\frac{(1+\tanh x)}{2}$.

Consider a bipartite density matrix $\rho_{AB}$. The weak measurement performed on the subsystem $B$ will yield the 
post-measurement state $\rho_{A|P^{B}(x)}$. Thus, we have 
\begin{align}
p(x)\rho_{A|P^{B}(x)} =  \mbox{Tr}_B[\rho_{AB}(I \otimes E(x))] 
= \sum_{i=0}^{1} a_i(x)p_{i}\rho_{A|i}.
\end{align}
Now the weak conditional entropy of the density operator is given by 
\begin{align}
& \sum_{y=x,-x}p(y)S(\rho_{A|P^{B}(y)})  =  \sum_{y=x,-x}p(y) S(\sum_{i=0}^{1} \frac{a_i(y)p_{i}}{p(y)}\rho_{A|i}) \nonumber\\
& \geq  \sum_{y=x,-x}p(y) \sum_{i=0}^{1} \frac{a_i(y)p_{i}}{p(y)}S(\rho_{A|i}) 
 = \sum_{i}p_{i}S(\rho_{A|i}). 
\end{align}
In the above we used concavity of the entropy. This then implies that 
\begin{align}
\label{weak}
S_w(A|\{P^{B}(x)\})\geq S(A|\{\Pi^B\}).
\end{align}
Thus, it shows that the weak conditional entropy is greater than the strong conditional
entropy for all possible measurement basis. Let $\{ {\tilde \Pi}_i^B \}$ be the measurement basis that minimizes the conditional entropy for the normal discord.
Then, from (\ref{dis}), we have $D(A:B) = S(A|\{ {\tilde \Pi}^B \} - S(A|B)$. Now define the weak measurement operators ${\tilde P}^B(\pm x) = 
\sqrt{a_0(\pm x)} {\tilde \Pi}_0^B  + \sqrt{a_1(\pm x)} {\tilde \Pi}_1^B $. Then, from (\ref{weak}), we have $S_w(A|\{ {\tilde P}^{B}(x)\}) \geq S(A|\{ {\tilde \Pi}^B\}$.
Hence, $S_w(A|\{ {\tilde P}^{B}(x)\}) - S(A|B) \geq S(A|\{ {\tilde \Pi}^B \} - S(A|B)$, i.e., 
the super quantum discord is always greater than the normal discord.

 Therefore, given a set of projectors that minimizes the conditional entropy for the normal discord, we can always define a set of 
weak measurement operators such that the super discord is larger than the normal discord. It may be noted that for the case of 
strong measurements, i.e., $\lim x \rightarrow \infty $, we have $D_w(A:B) = D(A:B)$ and thus, the equality holds. The extra 
quantum correlation revealed by the super quantum discord is possible only with the weak measurements. When we disturb the 
subsystem of a composite system weakly, we are destroying less quantumness. However, the strong measurement process tends to 
loose the extra quantumness irrevocably, and converts this to classical correlation. Hence, it is not captured by the normal quantum discord.
Thus, our result provides a new way of thinking about the quantum correlation.
The quantum correlation in a composite system does not have an absolute value, rather it is a relative notion. It
depends on the observer and on the strength of the measurement procedure. In a recent result it is argued that the
quantum mutual information can behave as if it is exclusively quantum \cite{syn,bennett}.


One should note that for the case $x=0$, there is no measurement performed on the system. Hence, 
the value of the super quantum discord is equal to the quantum mutual information. One can also 
prove that the super quantum discord always
decreases with increasing $x$, and thus covers all values between the
mutual information and the standard quantum discord.

\noindent
{\bf Theorem 2}: The super quantum discord is a monotonically decreasing function of the measurement strength, i.e., $\forall$
 $( x, y) \in [0, \infty] $ such that $x \geq y$, then $D_w(x) \leq D_w(y)$.

\noindent
{\bf Proof:} Proving $\frac{\partial D_w(x)}{\partial x} \leq 0$ will suffice the proof of above theorem. In (\ref{def}), without loss of generality
we assume that the super quantum discord is obtained for measurement basis $\{\Pi_i\}$. Now, we have 
\begin{align}
\label{r1}
 &\frac{\partial D_w(x)}{\partial x} = \frac{\partial S_w(A|\{P^{B}(x)\})}{\partial x}\nonumber \\
 &= \frac{\partial p(x)}{\partial x} S(\rho_{A|P^{B}(x)}) + p(x)\frac{\partial}{\partial x} S(\rho_{A|P^{B}(x)})\nonumber \\
 &+\frac{\partial p(-x)}{\partial x}S(\rho_{A|P^{B}(-x)}) + p(-x) \frac{\partial}{\partial x}S(\rho_{A|P^{B}(-x)}).
\end{align}

Using 
\begin{align}
\label{r4}
 &\frac{\partial}{\partial x} S(\rho_{A|P^{B}(\pm x)}) \nonumber \\
&=-\frac{1}{p(\pm x)} \Big\{\mbox{Tr}_A[\ln(\rho_{A|P^{B}(\pm x)})\frac{\partial}{\partial x}(p(\pm x)\rho_{A|P^{B}(\pm x)})]\nonumber \\
&~~~~~~~~~~~~~~~~~~~~~~+ \frac{\partial p(\pm x)}{\partial x}S(\rho_{A|P^{B}(\pm x)})\Big\}
\end{align}
we can express 
\begin{align}
\label{r6}
&\frac{\partial D_w(x)}{\partial x}
=-\mbox{Tr}_A \Big\{\ln(\rho_{A|P^{B}(x)})\frac{\partial}{\partial x}\mbox{Tr}_B[P^{B}(x) \rho_{AB} P^{B}(x)]\nonumber \\
 &~~~~~+ \ln(\rho_{A|P^{B}(-x)})\frac{\partial}{\partial x}\mbox{Tr}_B[P^{B}(-x) \rho_{AB} P^{B}(-x)]\Big\}.
\end{align}
This can be rewritten as 
\begin{align}
\label{r11}
\frac{\partial D_w(x)}{\partial x}
 &=-\frac{1}{2\cosh^2x}\mbox{Tr}_A\Big\{p_\psi \rho_{A|\psi}[\ln(\rho_{A|P^{B}(-x)})-\ln(\rho_{A|P^{B}(x)})] \nonumber \\
 &~~~~~+ p_{\bar\psi} \rho_{A|{\bar\psi}}[ \ln(\rho_{A|P^{B}(x)})- \ln(\rho_{A|P^{B}(-x)})]\Big\}\nonumber \\
 &=-\frac{1}{2\cosh^2x}R(x),
\end{align}
where $R(x) = \mbox{Tr}_A\Big\{p_\psi \rho_{A|\psi}[\ln(\rho_{A|P^{B}(-x)})-\ln(\rho_{A|P^{B}(x)})]+ p_{\bar\psi} \rho_{A|{\bar\psi}}[ \ln(\rho_{A|P^{B}(x)})- \ln(\rho_{A|P^{B}(-x)})]\Big\}$.
Now consider the state
\begin{align}
\label{r12}
\rho_{A|P^{B}(\pm x)} &= \frac{1}{p(\pm x)}\mbox{Tr}_B [P(\pm x)\rho_{AB}P(\pm x)] \nonumber \\
&= \frac{a^2(\pm x)p_\psi}{p(\pm x)} \rho_{A|\psi}+ \frac{a^2(\mp x)p_{\bar\psi}}{p(\pm x)}\rho_{A|{\bar\psi}},
\end{align}
where $p(\pm x) =a^2(\pm x)p_\psi+a^2(\mp x)p_{\bar\psi}$. Define $q(\pm x) = \frac{a^2(\pm x)p_\psi}{p(\pm x)}$, 
where $q(\pm x) \in [0,1]$, to write
\begin{align}
\label{r13}
\rho_{A|P^{B}(\pm x)} &= q(\pm x) \rho_{A|\psi}+ (1-q(\pm x))\rho_{A|{\bar\psi}}.
\end{align}
We can revert the above equations to obtain 
\begin{align}
\label{r14}
&\rho_{A|\psi} = -k(x) \rho_{A|P^{B}(x)} + l(x) \rho_{A|P^{B}(-x)},\\
&\rho_{A|\bar\psi} = r(x) \rho_{A|P^{B}(x)} - s(x) \rho_{A|P^{B}(-x)},
\end{align}
where $k(x) = (1-q(-x))/u(x) $, $l(x) = (1-q(x))/u(x)$, $r(x) = q(-x)/u(x)$ and $s(x) = q(x)/u(x)$ with $u(x) = q(-x) - q(x)$.
\begin{align}
\label{r15}
u(x) &= \frac{a^2(x)p_\psi}{p(x)} - \frac{a^2(-x)p_\psi}{p(-x)}
=\frac{p_\psi \tanh x}{p(x)p(-x)}.
\end{align}
Thus $u(x) \geq 0$ for $x\in[0,\infty]$ and this in turn shows that $k(x)$, $l(x)$, $r(x)$, $s(x)$ all
are positive semidefinite. Therefore, we can express $R(x)$ as 
\begin{align}
\label{r16}
R(x) 
&=[p_\psi k(x) + p_{\bar\psi} r(x)] S(\rho_{A|P^{B}(x)}||\rho_{A|P^{B}(-x)}) \nonumber \\
&+ [p_\psi l(x) + p_{\bar\psi} s(x)]S(\rho_{A|P^{B}(-x)}||\rho_{A|P^{B}(x)})\nonumber \\
&\geq 0,
\end{align}
where $S(\rho|| \sigma) =  \mbox{Tr} (\rho \log \rho - \rho \log \sigma) $ 
is the relative entropy and is always greater than or equal to zero. Now from (\ref{r11})
we conclude that $\frac{\partial D_w(x)}{\partial x} \leq 0$. This completes the proof of the theorem.

In what follows, we provide some illustrative examples where the super quantum discord reveals 
more quantum correlation compared to the normal discord.

\emph{Super discord for two qubit pure entangled state.--}
Now consider a general pure entangled state of two qubits. In the Schmidt decomposition form this can be written as
$|\psi\rangle_{AB}=\sqrt{\lambda_0}|00\rangle+\sqrt{\lambda_1}|11\rangle$,
where $\lambda_0, \lambda_1$ are the Schmidt coefficients. It is known that the normal discord for a pure entangled state is equal to its 
entanglement entropy, i.e., $D(A:B) = S(\rho_A) = S(\rho_B)$. However, 
this is not so for the super quantum discord. This is a new feature of the weak measurement. We will show that the super quantum discord can exceed the quantum 
entanglement for any pure bipartite state. Let us now define the weak operator in an arbitrary measurement basis as
$P^{B}(x) = a(x) \Pi_{\psi}   + a(-x) \Pi_{{\bar \psi}}$ ,
where  $a(\pm x)= \sqrt{\frac{(1 \mp \tanh x)}{2}}$ and  $\{ \ket{\psi(\theta,\phi)}, \ket{\bar{\psi}(\theta,\phi)}  \}$ are arbitrary 
single qubit basis.
Similarly we can define $P^{B}(-x)$. The action of the weak operator $P^{B}(x)$ on the pure state results in 
 $\rho_{A|P^{B}(x)}  =\frac{1}{2 p(x)}[ \lambda_0 (1-\tanh x\cos\theta)|0\rangle\langle0| 
  + \lambda_1 (1+\tanh x\cos\theta)|1\rangle\langle1| 
 - \sqrt{\lambda_0\lambda_1} \sin\theta \tanh x (e^{i\phi} |0\rangle\langle1| + e^{-i\phi}|1\rangle\langle0|) ]$. 
The probability of outcome corresponding to measurement operator $P^{B}(x)$ is given by
 $p(x) = \frac{1}{2}{[1-(\lambda_0-\lambda_1)\tanh x \cos \theta]}$.
The super quantum discord for the pure entangled state is 
given by
\begin{align}
& D_w(A:B) =  -\lambda_0\log\lambda_0 -\lambda_1\log\lambda_1\nonumber \\
&-   \min_{\theta} \sum_{y=\pm x} p(y) [ k_+(y)\log k_+(y) + k_-(y) \log k_-(y)],
\end{align}
where $k_{\pm}(y)  =\frac{1}{2}[1\pm \sqrt{1 - \frac{\lambda_0\lambda_1}{(p(y)^2\cosh^2{y})}}]$.

Now in the strong measurement limit ( i.e. $x\rightarrow \infty$) the eigenvalues of measured density matrix are $0$ and $1$, independent of 
the basis parameters $\theta$ and $\phi$. This implies that $D_w(A:B)=-\lambda_0\log\lambda_0 -\lambda_1\log\lambda_1$, and it is equal to the normal discord.
For maximally entangled pure state case 
the super discord is $D_w(A:B)=1-[\frac{1-\tanh x}{2}\log{\frac{1-\tanh x}{2}} + \frac{1+\tanh x}{2}\log{\frac{1+\tanh x}{2}}]$.
In the weak measurement limit ( very small $x$), $D_w(A:B)=1-[\frac{1-x}{2}\log{\frac{1-x}{2}} + \frac{1+ x}{2}\log{\frac{1+ x}{2}}]$ and it is always greater than one.
For example, when $x=0.2$, $D_w(A:B)=1.4689 > D(A:B)$. Therefore, for the weak measurement, the super discord is greater than the normal discord 
(entanglement entropy), i.e., $D_w(A:B)> D(A:B)=S(B)$.


\emph{Super discord for the Werner state.---}
The Werner state is a prime example to show that the quantum discord is nonzero in the separable regime. Now, we will show that the quantum correlation revealed by the 
super discord for the Werner state is also more than that of the normal discord. 
The Werner state is an admixture of a random state and a maximally entangled state, namely,
\begin{equation}
\rho_{AB} = z |\Psi^-\rangle\langle\Psi^-|+\frac{(1-z)}{4}I,
\end{equation}
where $|\Psi^-\rangle=(|01\rangle - |10\rangle)/\sqrt{2}$.
The corresponding entropies for the above state are given by 
$ S(B)=1$,
 $ S(A,B)=-\frac{3(1-z)}{4}\log(\frac{1-z}{4})-\frac{(1+3z)}{4}\log(\frac{1+3z}{4})$ and 
$ S(A|\{\Pi^B\})=-\frac{(1-z)}{2}\log(\frac{1-z}{2})-\frac{(1+z)}{2}\log(\frac{1+z}{2})$.
In defining the conditional entropy, we have used the computational basis. Since the Werner state is rotationally invariant therefore 
this yields the same result for the conditional entropy for any 
measurement basis and hence we do not have to minimize it over all measurement bases. 

If we perform the weak measurement on the subsystem $B$, then for the Werner state 
the SQD is given by
\begin{align}
&D_w(A:B) =\frac{3(1-z)}{4}\log(\frac{1-z}{4})+\frac{(1+3z)}{4}\log(\frac{1+3z}{4})\nonumber\\
&+1-\sum_{y=x,-x}\frac{(1-z\tanh y)}{2}\log(\frac{1-z\tanh y}{2}). 
\end{align}
This is larger than the normal quantum discord. 
In the strong measurement limit $(x\rightarrow\infty)$, the last term in the above expression becomes the strong conditional entropy and 
hence $\lim_{x\rightarrow\infty}D_w(A:B)=D(A:B)$.

\begin{figure}[htbp]
\centering
{
\includegraphics[width=60 mm]{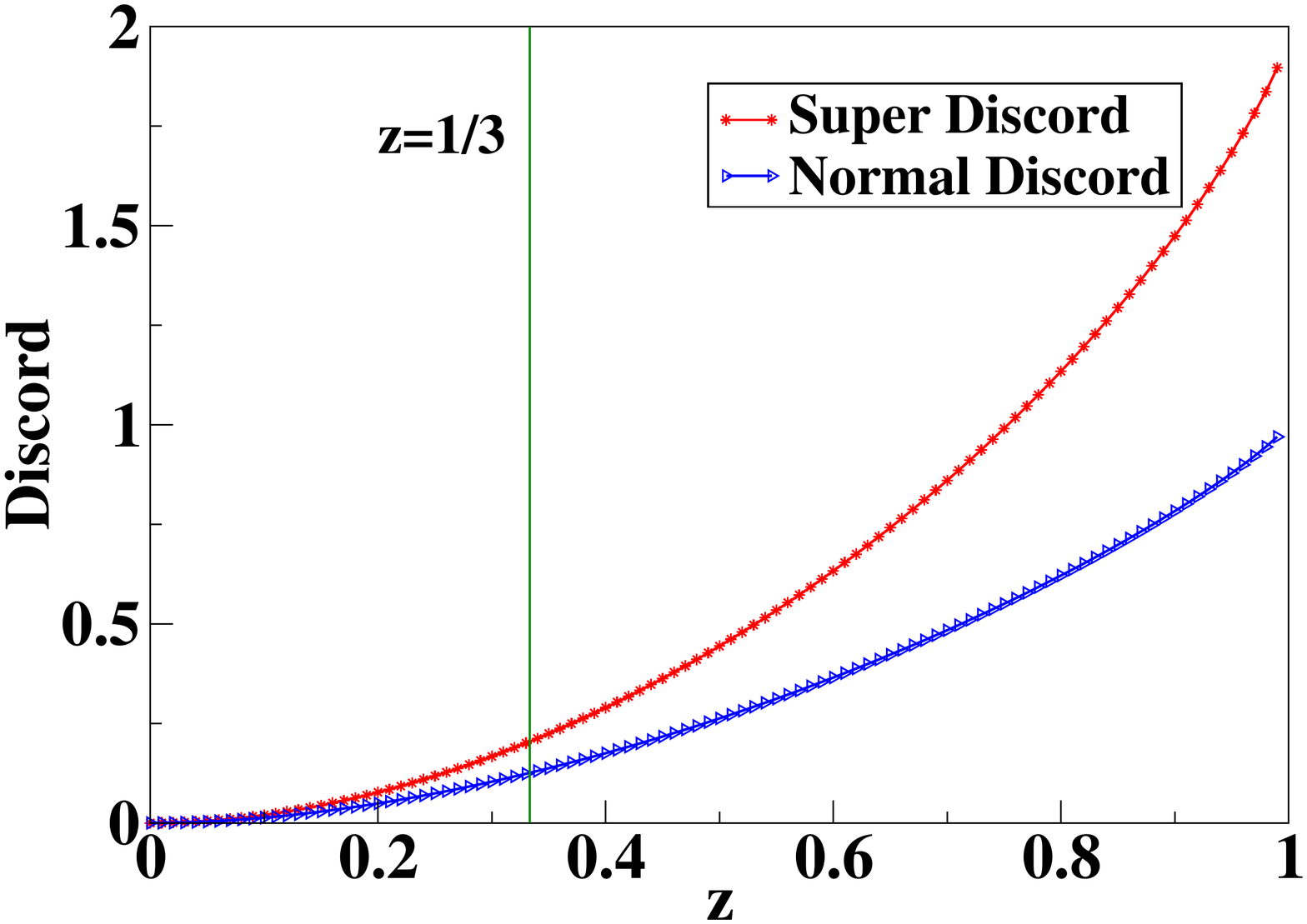}
}
\caption{(Color online) The super and the normal discords as a function of $z$ for the Werner state at $x=0.2$. It shows that the super quantum discord is 
greater than the normal discord for all $z$.}

\end{figure}



\emph{ Weak discord for two-qubit density operator:--}
Consider a density matrix which is locally unitarily equivalent to the most general density matrix \cite{luo} 
\begin{align}
 \rho_{AB} &= \frac{1}{4}[I +(  \mbox{\bf a} \cdot \sigma \otimes I)+ (I \otimes \mbox{\bf b} \cdot  \sigma ) \nonumber \\
& + \sum_{i=1}^{3} c_i(\sigma_i \otimes \sigma_i)],
\end{align}
where $\mbox{\bf a}$, $\mbox{\bf b}$ are the three dimensional real vectors and $c_i$ are real. The weak conditional entropy for this state 
is given by
\begin{align}
&S_w(A|\{P^{B}(x)\}) = -p(x)[\lambda_+(x)\log\lambda_+(x)\nonumber \\
&+\lambda_-(x)\log\lambda_-(x)] -p(-x)[\lambda_+(-x)\log\lambda_+(-x)\nonumber \\
&+\lambda_-(-x)\log\lambda_-(-x)],
\end{align}
where
\begin{align}
&p(x)=[1-( \mbox{\bf b} \cdot \mbox{\bf n}) \tanh x]/2,\nonumber \\
&\lambda_\pm(x) = \frac{[1-( \mbox{\bf b} \cdot \mbox{\bf n})\tanh x]\pm\sqrt{\sum_{i}(a_i - c_i n_i \tanh x)^2}}{2[1- ( \mbox{\bf b} \cdot \mbox{\bf n})\tanh x]} \nonumber \\
& \mbox{\bf n} = (\sin\theta\cos\phi,\sin\theta\sin\phi,\cos\theta).
\end{align}


Now Consider an arbitrary density matrix with $a_1=0.01$, $a_2=0.1$, $a_3=0.22$, $b_1=0.1$, $b_2=0.03$, $b_3=0.5$, $c_1=0.1$, $c_2=0.02$, $c_3=0.2$, in the Bloch normal form.
For this state $S(B)=0.80262$ and $S(AB)=1.732$. The super quantum discord is given by $D_w{(A:B)}=S_w(A|\{P^{B}(x)\})-0.92938$. 
Using the expression for the weak conditional entropy, we have plotted the super discord as
a function of $x$, $\theta$ at a fixed value of $\phi$. This shows that indeed the SQD is larger than 
the normal quantum discord.
\begin{figure}[htbp]
\centering
{
\includegraphics[width=85 mm]{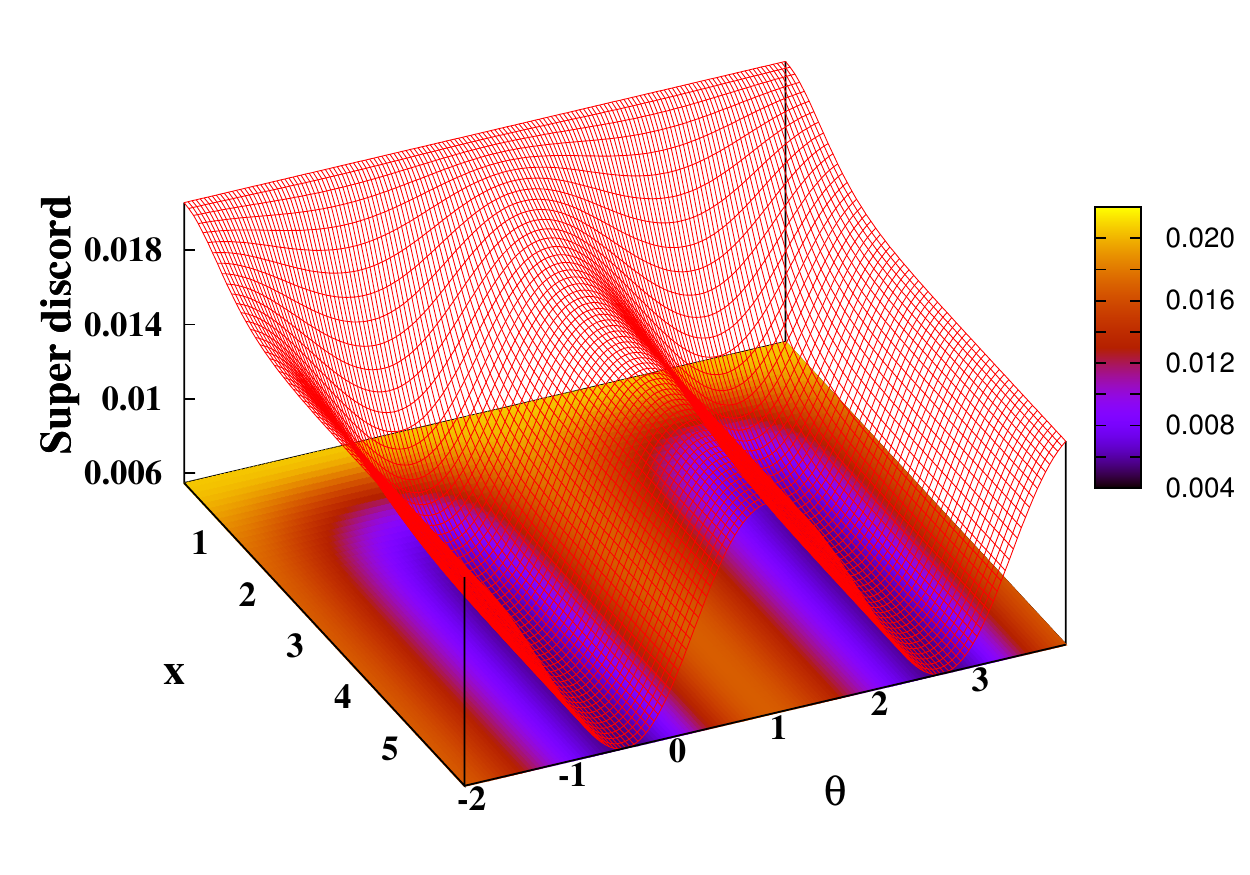}
}

\caption{(Color online) Weak discord for above considered density matrix as a function of $\theta$ and $x$ at fixed $\phi =1.57$. At $\theta =0$, for larger $x$ ($\approx x>3$) weak discord is constant and is equal
to normal discord at fixed $\theta= 0$ and $\phi =1.57$. Also for smaller values of $x$ ($\approx x<3$) weak discord is greater than its value at larger $x$ (normal discord). }
\end{figure}

\emph{Conclusion:--}
To conclude, here we have raised a fundamental question: Can weak measurements reveal more quantumness? We have shown that indeed the 
answer to this question is in affirmative. We have proved that the weak measurement can capture more quantum correlation of a 
bipartite system than the strong measurement. This gives rise to the notion of super quantum discord that goes beyond 
the normal quantum discord.  We have proved the monotonicity of the super quantum discord, i.e., it always
decreases with the increase of measurement strength, and thus covers all values between the
mutual information and the standard quantum discord.
The super discord has been calculated explicitly for the pure entangled state and the Werner state.
For pure entangled state the super quantum discord can be more than the quantum entanglement. 
As the quantum discord depends on the observer
who is trying to access the quantum system by performing measurement on one of the subsystems, 
it is an observer dependent notion. Our result shows that quantum
correlation is not only an observer dependent notion but also depends on how weakly one perturbs the system. If we perform
quantum measurement weakly, then there can be more quantum correlation between the subsystems which can be exploited for 
practical use. Thus, the notion of super discord can be potentially a useful resource for quantum computation,
quantum communication and general quantum information processing tasks. In future, it will be worth exploring if the weak 
measurement can enhance the entanglement assisted classical capacity and coherent information. 
This will open up new avenues of
explorations in quantum technology with weak measurements.

\end{document}